\documentclass{fundam}

\publyear{22}
\papernumber{2102}
\volume{185}
\issue{1}

\usepackage{url} % takes care of hyperlinks, preferred over hyperref
\usepackage[ruled,lined]{algorithm2e}% provides Algorithm environment
\usepackage{graphicx}% allows for inclusion of graphic files (figures)
\usepackage{tikz}
\usetikzlibrary{automata, positioning, arrows}
\usepackage{amsmath}

\begin{document}

\title{Embedding into Special Classes of Cyclic Graphs and its Applications in VLSI Layout}
\address{$^*$Department of Mathematics, Hindustan Institute of Technology and Science,\\ Chennai, India, 603103, vprsundar{@}gmail.com}

\author{R. Sundara Rajan$^*$, \textbf{Rini Dominic D.} \\
Department of Mathematics, \\Hindustan Institute of Technology and Science,\\ Chennai, India, 603103\\
\and T.M. Rajalaxmi \\
Department of Mathematics, \\Sri Sivasubramaniya Nadar College of Engineering,\\ Chennai, India, 603103\\
\and L. Packiaraj \\
Department of Mathematics, \\J.P. College of Arts and Science, \\Tenkasi, India, 627852\\
} 

\maketitle

\runninghead{R. Sundara Rajan et al.} {Embedding into Special Classes of Cyclic Graphs and its Applications in VLSI Layout}

\begin{abstract}
  Graph embedding is the major technique which is used to map guest graph into host graph. In architecture simulation, graph embedding is said to be one of the strongest application  for the execution of parallel algorithm and simulation of various interconnection networks \cite{Pa99}. In this paper, we have embedded circulant networks into star of cycle and folded hypercube into cycle-of-ladders and compute its exact wirelength. Further we have discussed the embedding parameters in VLSI Layout.
\end{abstract}


\section{Introduction}

Recent study in integrated circuit has made the construction of interconnection network possible. Along with these studies, most of the interconnection typologies has been investigated and discussed. Most often the interconnection networks are been designed as graphs. Graph embedding is said to be one of the major factor that is used to evaluate these interconnection networks. It can also be used for many parallel computing applications, like network design, data structures and so on. The evaluation of embedding can be done by using different cost criteria. The main criteria is the wirelength. The embedding performance is also measured by dilation, edge congestion, load, and expansion.\\
An embedding of $A$ into $B$ is the pair $(g,P_g)$ described {\rm \cite{ChSt95}} as follows:
\begin{enumerate}

\item $g$ is the $1 - 1$ mapping from $V(A)$ to $V(B)$
\item $P_g$ is the $1 - 1$ mapping from $E(A)$ to $P_g(e):P_g (e)$ be the path in host graph joining $g(x)$ and $g(y)$ for $e = xy \in E(A)$.
\end{enumerate}

Let $g$ be the embedding from $A$ into $B$. The \textit{edge congestion} for $e \in E(B)$, denoted by $EC_g(e)$, is the cardinality of edges $(uv)$ in $A$ for which $e$ is in the path $P_g(uv)$ joining $g(u)$ and $g(v)$ in $B$
Specifically,
\begin{equation*}
EC_{g}(A,B(e))=\left\vert \left\{ (x,y)\in E(A):e\in
P_{g}(x,y)\right\} \right\vert
\end{equation*}
where $P_{g}(x,y)$ represents the path between $g(x)$ and $g(y)$ in $B$
with respect to $g$.

The wirelength of embedding $g$
of $A$ to $B$ is \cite{ChSt95} given by
\begin{equation*}
WL_{g}(A,B)=\underset{(x,y)\in E(A)}{\sum }d_{B}(g(x),g(y))=\underset{e\in
E(B)}{\sum }EC_{g}(A,B(e))
\end{equation*}
where\textit{\ }$d_{B}(g(x),g(y))$ denotes the length of the path $%
P_{g}(x,y)$ in $B$.
Then, the \textit{
wirelength} of $A$ into $B$ is explained as
\begin{equation*}
WL(A,B)=\min WL_{g}(A,B)
\end{equation*}
here the minimum is taken from all the embedding $g$ of $A$ into $B$.

Graph embedding has been studies for enhanced hypercube into windmill and necklace graphs \cite{JbMaJn19}, hypercubes to necklace graphs \cite{IrBr12}, complete multipartite graphs into path, cycle, wheel, circulant graphs \cite{RsTr20}. The exact wirelength has been obtained for $1$-fault hamiltonian graphs into wheels and fans \cite{MaPa11}. The sharp lower bound has been determined for dilation as well as congestion of the networks such as wheel, fan, windmill, friendship graph into hypertrees \cite{RsTr19}. An algorithm has been determined with minimum wirelength for embedding Tur\'{a}n graphs into incomplete hypercubes \cite{AaSa21}, binary trees into hypercubes \cite{ChSt95}, complete binary trees into hypercubes \cite{Be01}, star graph to path \cite{Ya09}, ternary tree into hypercube \cite{PMa10}, meshes into m\"obius cubes
\cite{Ts08}, folded hypercube into grids \cite{BeCh98}.

In this paper, we  have obtained an algorithm of embedding circulant networks into star of cycle and folded hypercube into cycle-of-ladders and by Modified Congestion Lemma \cite{MaRaRaMe09,MiRaPaRa15} and Partition Lemma \cite{MaNa20} obtained the exact wirelength.

\section{Preliminaries}

In this section we have discussed the fundamental definitions and Lemma that are associated to embedding problems.

\paragraph{\rm \textbf{Edge isoperimetric problem:}} Let $S$ be a subset of $V(A)$, then
\begin{itemize} 
    \item $I_A(S)=\{(xy) \in E ~|~ x, y \in S\},~   I_A({a})=\max\limits_{S \subseteq V, |S|=a} |I_A(S)|$
    \item $\theta_A(S)=\{(xy) \in E ~|~ x \in S, y \not\in S\},~ \theta_A(a)=\min\limits_{S \subseteq V, |S|=a}|\theta_A(S)|.$
\end{itemize}

For $a=1,2, \ldots, n$, the problem that is used for obtaining the subset $S$ of $V(A)$ such that $|S|=a$ and $|I_A(S)|=I_A(a)$ is said to be the \textit{Maximum Subgraph Problem} (MSP) \cite{RsTr20}. Also, for $a=1,2, \ldots, n$, the problem which is used to obtain a subset $S$ of $A$ with $|S|=a$ and $|\theta_A(S)|=\theta_A(a)$ is known as the \textit{Minimum Cut Problem} (MCP) \cite{RsTr20}.
In the case of regular graph, $I_A$ and $\theta_A$ is similar in such a way that a answer for one gives the answer for the other. In another words, for any regular graph with regularity say $r$, $2I_A(a)+\theta_A(a)=ra, a=1,2, \ldots, |V(A)|$.

\begin{lemma} {\rm (Modified Congestion Lemma)} {\rm\cite{MaRaRaMe09,MiRaPaRa15}}
\label{lemmacongestion1}Let $g$ be the embedding from $A$ into the host graph $B$ having $|V(A)|=|V(B)|$. Let the edge cut of $B$ be $T$ that is by removing $T$, we can seperate $B$ exactly into two connected components $B_{1}$ and $B_{2}$, and $A_{1} =G[g^{-1}(V(B_{1}))]$ and $A_{2} = G[g^{-1}(V(B_{2}))]$. Further, let $T$ satisfies the below conditions:
\begin{itemize}
\item [$(i)$] For each $xy\in E(A_{j})$, {$j\in \{1,2\}$,} $P_{g}(xy)$ have no edges in $T$.
\item [$(ii)$] For each $xy\in E(A)$, $x\in V(A_{1})$, $y\in V(A_{2})$, $P_{g}(xy)$ have only one edge in $T$.
 \item [$(iii)$] $V(A_{1})$ and $V(A_{2})$ are the optimal sets with the cardinality of $V(A_{1})$ and $V(A_{2})$ respectively.
\end{itemize}

If $T$ exists, then $EC_{g}(T)$ is the minimum over all the embeddings $g:A \rightarrow B$ and
\begin{equation*}
   EC_{g}(T)=\underset{u\in V(A_1)}{\sum } \deg_A(u)-2|E(A_1)|=\underset{u\in V(A_2)}{\sum } \deg_A(u)-2|E(A_2)|\,.
\end{equation*}

\end{lemma}

\begin{lemma} {\rm (Partition Lemma)}  {\rm\cite{MaNa20}}
\label{partitionlemma} Consider an embedding $g$ of $A$ into $B$. Let $E(B)$ has the partition $\{X_{1}, X_{2}, \ldots, X_{l}\}$ where each $X_{i}$ is the edge cut of host graph and satisfied the conditions of Lemma~\rm\ref{lemmacongestion1}, then
\begin{equation*}
WL_{g}(A,B)=\overset{l}{\underset{i=1}{\sum }}EC_{g}(X_{i}).
\end{equation*}
Moreover, $WL(A,B)=WL_{g}(A,B)$.
\end{lemma}

\section{Embedding folded hypercubes into cycle-of-ladders}

In this section we embed folded hypercubes cycle-of-ladders. We need the following to move further.

Among the interconnection networks, the most efficient and popular topological structure is the hypercube. In parallel processing hypercube has become the first choice since it has many exceptional features. 

Fang {\rm \cite{JyFa08}} has put forward fairly a new cycle-embedding feature known as the bipancycle connectivity and also a new graph known as the cycle-of-ladders. By giving algorithms for embedding $COL(l,r)$ into $Q^{s}$ and by bringing out the bipanconnected cycles in the $COL(l,r)$, the hypercube $Q^{s}$ have been proved to be a bipancycle connected graph.

$P(r)$ and $C(r)$ represents a path and cycle  having length $r-1$ and $r$ respectively. A ladder with length $r$, represented by $L(r)$, is a $P(r) \times K_2$ (the cartesian product $A \times B$ of graphs $A$ and $B$ with disjoint vertex set $V_1$ and $V_2$ and edge set $E_1$ and $E_2$ is the graph with vertex set $V_1 \times V_2$ and $a=(a_1, a_2)$ adjacent with $b=(b_1, b_2)$ whenever $a_1=b_1$ and $a_2$ is adjacent to $b_2$ or $a_2=b_2$ and $a_1$ is adjacent to $b_1$). Every vertex of an $L(r)$ is labelled as $(a_0, a_1)$, where $a_0=0$ or $a_0=1$, and $0 \leq a_1 \leq r$. Every edge $((0,a_1),(1,a_1))$ is
said to be a rung of $L(r)$, where $0\leq a_1\leq r$ and it is known as the $a_1^{th}$ rung. The bottom rung of the ladder is the $0^{th}$ rung. The bands of the $L(r)$ is the two paths $((0,0),(0,1),\ldots,(0,r))$ and $((1,0),(1,1),\ldots,(1,r))$ said to be the $0^{th}$ band and $1^{st}$ band respectively. Clearly $L(r)$ has $2(r+1)$ vertices and $3r+1$ edges.
\begin{definition} \label{definition}{\rm\cite{JyFa08}}
A graph obtained by unifying a bone cycle $B_c$ and $l$ ladders $L_d(0), L_d(1),\ldots,L_d(l - 1)$ with the bottom rungs $B_r(0),B_r(1),\ldots,B_r(l - 1)$, respectively, such that every $B_r(i)$ is in the $B_c$ where $0 \leq i \leq l - 1$ is called cycle-of-ladders.
\end{definition}

In this section, we consider that each $L_d(i)$ $0\leq i\leq l-1$, has length $r$. Clearly the cycle-of-ladders contains $2l(r+1)$ vertices and it is represented by $COL(l,r)$.

\begin{theorem}
\label{TheoremTTT0} {\rm \cite{HsNi97, Ha04, BoGuSh94}} Let $Q^{s}$ be an $s$%
-dimensional hypercube. For $1\leq j\leq 2^{s}$, $L_{j}$ is an optimal set on $j$
vertices.
%using lexicographic labeling \cite{BeChHaRoSc98}.
\end{theorem}

\begin{definition}{\rm \cite{Xu01}}
For two vertices $u = u_{1}u_{2}\cdots u_{s}$ and $v = v_{1} v_{2}\cdots v_{s}$ of $Q^{s}$, $(u, v)$ is a complementary edge if
and only if the bits of $u$ and $v$ are complements of each other, that is, $v_{j} = \overline{u}_{j}$ for each $j = 1, 2, \ldots , s$. The $s$-dimensional
folded hypercube, denoted by $FQ^{s}$ is an undirected graph derived from $Q^{s}$ by adding all complementary edges.

\end{definition}

\begin{lemma}{\rm \cite{RaAr11}}
Let $FQ^{s}$ be an $s$-dimensional folded hypercube. Let the subgraph of $FQ^{s}$ that is  isomorphic to $L_{p}$ be $P$ where $p \leq 2^{s-1}.$ If $P_{1}$ and $P_{2}$ are vertex disjoint subgraphs of $FQ^{s}$ given by $p_{1}$ and $p_{2}$ successive vertices of $0Q^{s-1}$
or $1Q^{s-1}$ such that $p_{1}+ p_{2} = p$, then $|E(FQ^{s} [P_{1} \cup P_{2}])| \leq |E(FQ^{s} [P])|$.
\end{lemma}

\begin{lemma}{\rm \cite{RaAr11}}
In the folded hypercube $FQ^{s}$, $L_{j}$ is isomorphic to $RL_{j}$ for $1 \leq   j \leq   2^{s}$.
\end{lemma}

\paragraph{Embedding Algorithm A} \label{EmbeddingAlgorithmA}

\paragraph{Input :}

The $s$-dimensional folded hypercube $FQ^s$ and a cycle-of-ladders $COL(4,2^{s-3}-1), s\geq3$.

\paragraph{Algorithm :}

Labelling of the vertices of $FQ^s$ is by Gray code labeling  \cite{ChSt95} from $0$ to $2^s-1$. Label the vertices of $COL(4,2^{s-3}-1)$ as follows: For $1 \leq j \leq 4$, label the $0^{th}$ band  vertices of $L_d(j)$ from bottom to top as $0,1,\ldots,2^{s-3}-1$ and  label the $1^{st}$ band vertices of $L_d(j)$ from top to bottom as $(2j-1) (2^{s-3}-1+1),(2j-1)(2^{s-3}-1+1)+1,\ldots,(2j-1) (2^{s-3}-1+1)+ 2^{s-3}-1$ and label the $0^{th}$ band vertices from bottom to top  as $(2j-2) (2^{s-3}-1+1),(2j-2) (2^{s-3}-1+1)+1,\ldots,(2j-1)(2^{s-3}-1+1)-1$.

\paragraph{Output :}

An embedding $g$ of $FQ^s$ into $COL(4,2^{s-3}-1)$ given by $g(x)=x$ with minimum
wirelength.

\paragraph{Proof of correctness :}

\begin{figure}
\centering
\includegraphics[width=9 cm]{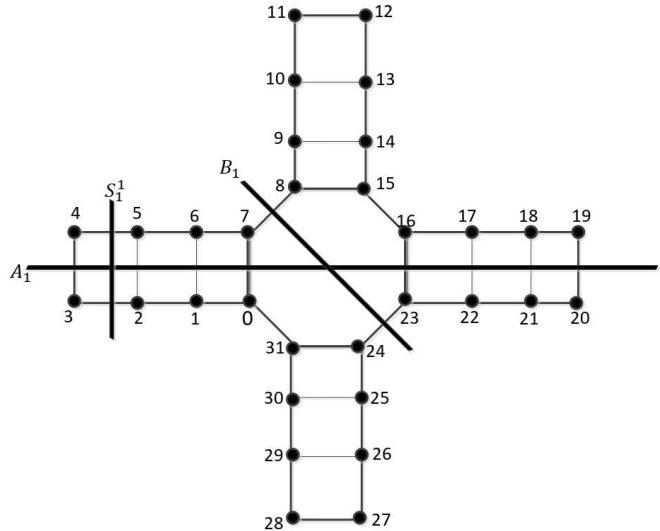}
%\vspace{-2 cm}
\caption{The edge cut of $COL(4, 3)$}
\label{fig1}
\end{figure}

For $1\leq i \leq 2$, let $A_i$ be the edge cut that consists of all the rungs of $L_d(i)$ and $L_d(2+i)$. For $1\leq j \leq 2$, let $B_j$ is the edge cut that has the edges between $L_d(i)$ and $L_d(i+1)$ and the edge between $L_d(2+i)$ and $L_d(2+i+1)$.
For $1\leq i \leq 4, 1\leq j \leq 2^{s-3}-1$, let $S_{i}^{j}$ be the edge cut in $L_d(i-1)$  which consists of the edges between $(2^{s-3}-1-j+1)^{th}$ rung and $(2^{s-3}-1-j)^{th}$ rung.
Then $\{A_i:1\leq i\leq 2\}\cup  \{ B_j:1\leq j \leq 2\} \cup  \{ S_{i}^{j}: 1\leq i\leq 4, 1\leq j\leq 2^{s-3}-1\}$ is a partition of $E(COL(k,s))$ (see Figure \ref{fig1}).

For each $i$, $1\leq i\leq 2$, $E(COL(k,r))\setminus A_i$ has two components $B_{i1}$ and $B_{i2}$ where
\begin{align*}
  V(B_{i1}) &= \{(2i-1)(2^{s-3}-1+1),(2i-1)(2^{s-3}-1+1)+1,\ldots,(2i-1)(2^{s-3}-1+1)\\
  &\quad+k(2^{s-3}-1+1)-1\}.
\end{align*}
Let $A_{i1}=G[V(g^{-1}(B_{i1}))]$
and $A_{i2}=G[V(g^{-1}(B_{i2}))]$. By Lemma \ref{lemmacongestion1},  $A_{i1}$, is optimal, and every $T_{i}$ satisfies all the conditions of the Modified Congestion Lemma. Therefore $EC_{g}(T_{i})$ is minimum.

For each $j$, $1\leq j \leq 2$, $E(COL(k,r))\setminus B_j$ has two components $B_{j1}$ and $B_{j2}$ where
\begin{align*}
  V(B_{j1}) &= \{(2j)(2^{s-3}-1+1),(2j)(2^{s-3}-1+1)+1,\ldots,(2j)(2^{s-3}-1+1)\\
  &\quad +k(2^{s-3}-1+1)-1\}.
\end{align*}
Let $A_{j1}=G[V(g^{-1}(B_{j1}))]$
and $A_{j2}=G[V(g^{-1}(B_{j2}))]$. By Lemma \ref{lemmacongestion1},  $A_{j1}$, is optimal, and each $X_{j}$ satisfies all the conditions of the Modified Congestion Lemma. Therefore $EC_{g}(X_{j})$ is minimum.

For each $i$, $j$, $1 \leq i \leq 4$, $1\leq j\leq 2^{s-3}-1$, $E(COL(k,r))\backslash S_{i}^{j}$ has two components $B_{i1}^j$ and $B_{i2}^j$ where

\begin{equation*}
V(B_{i1}^j)=\left\{
\begin{array}{lcl}
\{0,1,\ldots,j-1\}\cup\{2i(2^{s-3}-1+1)-1,2i(2^{s-3}-1+1)-2,\ldots,\\2i(2^{s-3}-1+1)-j\}  \hfill;\quad if\quad i=1 \\
\\
\{(2i-1)(2^{s-3}-1+1)-1,(2i-1)(2^{s-3}-1+1)-2,\ldots,\\(2i-1)(2^{s-3}-1+1)-j\}\cup \{(2i-1)(2^{s-3}-1+1),\\(2i-1)(2^{s-3}-1+1)+1,\ldots,(2i-1)(2^{s-3}-1+1)+j-1\} .
\end{array}%
\right.
\end{equation*}
Let $A_{i1}^j=G[V(g^{-1}(B_{i1}^j))]$
and $A_{i2}^j=G[V(g^{-1}(B_{i2}^j))]$. By Lemma \ref{lemmacongestion1} $A_{i1}^j$ is an optimal set, each  $T_{i}^{j}$ satisfies all the three conditions of the Modified Congestion Lemma. Therefore $EC_{g} (U_{i}^{j})$ is minimum. The Partition Lemma gives that the wirelength is minimum.
\begin{theorem} \label{firstthm}
Let $A$ be the $s$-dimensional folded hypercube $FQ^s$ and $B$ be the cycle-of-ladders $COL(4, 2^{s-3}-1)$. Then the wirelength of embedding from $A$ into $B$ is given by
$$WL(A,B)=2^{s+2}+4\overset{2^{s-3}-1}{\underset{j=1}{\sum }}\theta
_{A}(2j)
$$
\noindent{\textit{Proof.}}
By Embedding Algorithm A and by Partition Lemma, we have 
\begin{eqnarray*}
%\begin{split}
WL(A,B) &=& \overset{2}{\underset{i=1}{\sum }}EC_g(A_i)+\overset{2}{\underset{j=1}{\sum }}EC_g(B_j)+\overset{4}{\underset{i=1}{\sum }}~~\overset{2^{s-3}-1}{\underset{j=1}{\sum }}EC_g(S_i^j)\\
       &=& \overset{2}{\underset{i=1}{\sum }}2^{s}+\overset{2}{\underset{j=1}{\sum }}2^{s}+\overset{4}{\underset{i=1}{\sum  }}~~\overset{2^{s-3}-1}{\underset{j=1}{\sum }}\theta
_{A}(2j)\\
       &=& 2^{s+2}+4\overset{2^{s-3}-1}{\underset{j=1}{\sum }}\theta
_{A}(2j).
%\end{split}
\end{eqnarray*}
\end{theorem}

\newpage

\section{Embedding of circulant networks into star of cycle}
To show the main results we need the following definitions and theorems.
\begin{definition}{\rm\cite{Xu01}}
 A circulant graph $G(n;\pm R)$, where $R\subseteq \{1,2,\ldots,\left\lfloor n/2\right\rfloor\},$  $n\geq 3$ is said to be a graph having the vertex set $V=\{0,1,\ldots,n-1\}$ and the edge set $E=\{(i,j):\left\vert j-i\right\vert \equiv s (mod~n),$
$s\in R\}$.
\end{definition}

\begin{definition} \label{def}{\rm\cite{Xu01}}
A graph attained by replacing every vertex of star $K_{1,n}$ by a graph having n vertices is said to be star of graph and it
is represented by $G^{\star}$. The graph G which replaced at the center of $K_{1,n}$ we call the central copy of $G^{\star}$.
\end{definition}

 \begin{theorem}{\rm\cite{RaAr11}}
\label{TheoremTTT1} A set of $l$ successive vertices of $G(n;\pm 1)$, $%
1\leq l\leq n$ gives a maximum subgraph of $G(n;\pm S)$, where $%
S=\{1,2,\ldots,j\}$, $1\leq j<\left\lfloor n/2\right\rfloor $, $n\geq 3$.
\end{theorem}

\begin{theorem}{\rm\cite{RaRaRa12}}
\label{TheoremTTT2} The number of edges in a maximum subgraph on $l$
vertices of $G(n;\pm S)$, $S=\{1,2,\ldots,j\}$, $1\leq j<\left\lfloor
n/2\right\rfloor $, $1\leq l\leq n$, $n\geq 3$ is given by
\begin{equation*}
\xi =\left\{
\begin{array}{lcl}
l(l-1)/2 & ; & l\leq j+1 \\
lj-j(j+1)/2 & ; & j+1<l\leq n-j \\
\frac{1}{2}\{(n-l)^{2}+(4j+1)k-(2j+1)n\} & ; & n-j<l\leq n.
\end{array}%
\right.
\end{equation*}
\end{theorem}

\begin{definition} \label{def}{\rm\cite{RaRaRa12}}
Let $v_{1}, v_{2}, \ldots, v_{n}$ be successive vertices of central copy of $C_{n}^{\star}$
and $u_{i1}, u_{i2}, \dots, u_{in}$ be successive vertices of cycles
$C_{n}^{i}, i = 1, 2, \dots, n$. Let $e_{i}$ be edge such that $ e_{i} = v_{i}u_{i-1i}, \forall i = 2, 3, \ldots, n$ and $e_{1} = v_{1}u_{n1}$, which join each cycle $C_{n}^{i}$ with the central copy of $C_{n}^{\star}$.

\end{definition}

The structure of a star of cycle graph is $C_{n}^{\star}(0), C_{n}^{\star}(1),\ldots, C_{n}^{\star}(k-1)$. It is clear that this type of star of cycle has $m(k+1)$ vertices and it is represented by $C_{k}^{\star}(m)$.

\paragraph{Embedding Algorithm B} \label{EmbeddingAlgorithmA}

\paragraph{Input :}

 A circulant graph $G(n;\pm \{1,2,\ldots,j\})$, $1\leq j < \left\lfloor n/2\right\rfloor$  and a star of cycle $C_{k}^{\star}(m)$.

\paragraph{Algorithm :}

Label the consecutive vertices of $G(n;\pm 1)$ in $G(n;\pm{\{1,2,\ldots,j\}})$, as $0,1,\ldots, n-1$ in the clockwise sense. Label the vertices of $C_{k}^{\star}(m)$ as follows: For $0 \leq i\leq m-1,$ label $B_c(i)$ as $i(k+1)$. For $ 0 \leq i\leq m-1$, Label $C(i)$ as $(i(k+1)+1,i(k+1)+2,\ldots,i(k+1)+m$).

\paragraph{Output :}

An embedding $g$ of $G(n;\pm{\{1,2,\ldots,j\}})$ into $C_{k}^{\star}(m)$ given by $g(x)=x$ having wirelength minimum.

\paragraph{Proof of correctness :}
Let us assume that the labels are representing the vertices to which they are assigned.

\begin{figure}
\centering
\includegraphics[width=9 cm]{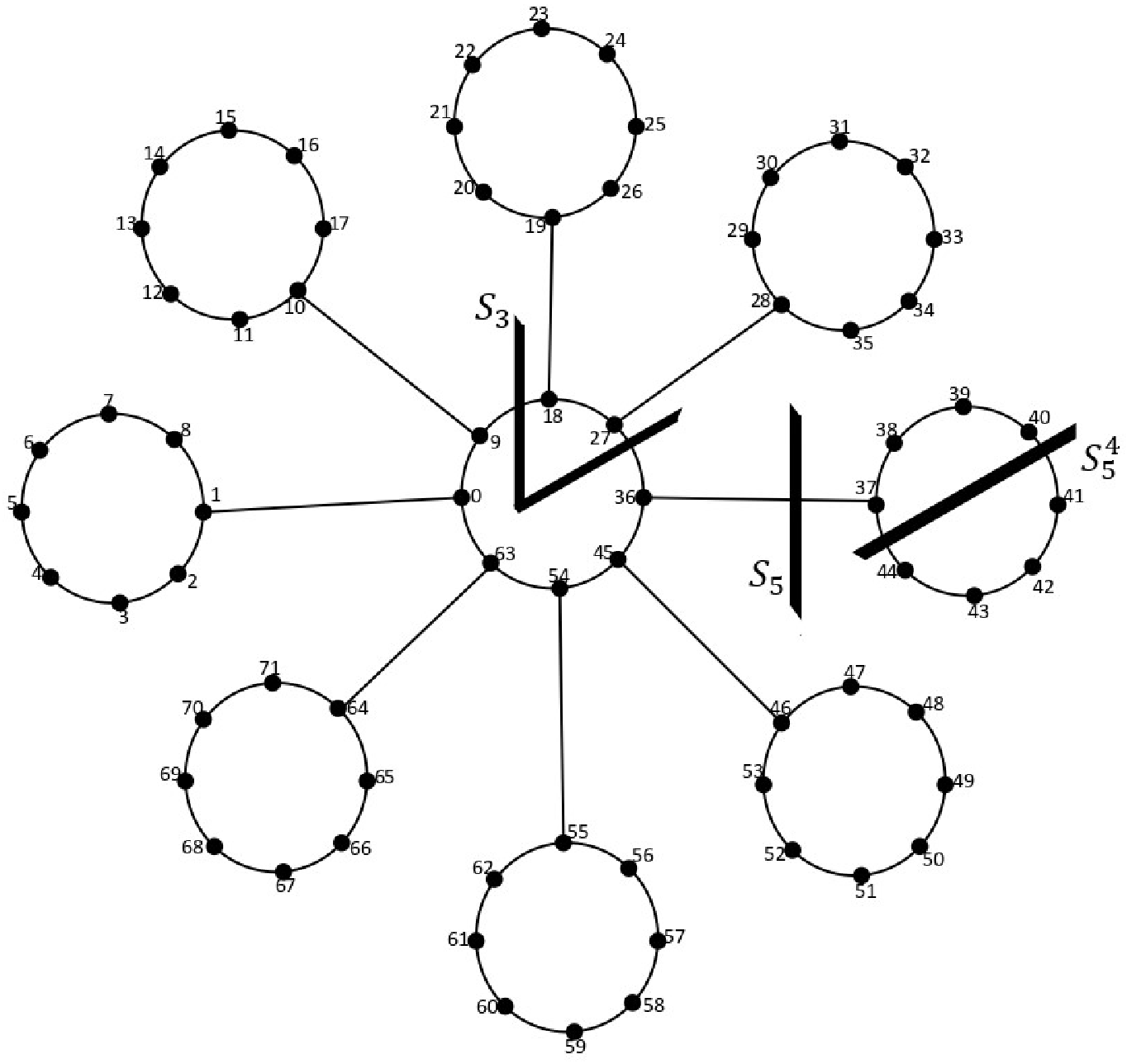}
%\vspace{-2 cm}
\caption{The egde cut of $EC_{8}^{\star}(8)$}
\label{fig3}
\end{figure}

\paragraph{\textbf{Case 1 ($m$ is even):}}
For $1\leq i \leq \frac{k}{2}$, let $S_i$ be the edge cut that has the edges between $C(i)$ and $C(i+1),$ together with the edges between $C(\frac{m}{2}+i)$ and $C(\frac{m}{2}+i+1({mod\ m))}$.
For $1 \leq j \leq \frac{k}{2},$ let $S_{j}$ be the edge cut that has edges between $C(j)$ and $B_c(j)$, For $1 \leq i \leq \frac{k}{2}$, $1\leq j \leq\frac{m}{2}$,  let $S_{i}^{j}$=$\{j(k+1)+i,(j(k+1)+(i+1)),((\frac{m}{2}+i)+(k+1)j,(\frac{m}{2}+i+1)+(k+1)j)\}$.
Then $\{S_i:1\leq i\leq \frac{k}{2}\}\cup  \{ S_j:1\leq j \leq \frac{k}{2}\} \cup  \{ S_{i}^{j}:1 \leq i \leq \frac{k}{2},1\leq j \leq\frac{m}{2}$ is a partition of $[E(C_{k}^{\star}(m))]$ (see Figure \ref{fig3}).

For $1\leq i\leq \frac{k}{2}$, $E(C_{k}^{\star}(m))\setminus S_i$ has two components $B_{i1}$ and $B_{i2}$, where
\begin{eqnarray*}
  V(B_{i1}) &=& \{i(k+1),i(k+1)+1,\ldots,i(k+1)+1+2k\}.
\end{eqnarray*}
Let $A_{i1}=G[V(g^{-1}(B_{i1}))]$
and $A_{i2}=G[V(g^{-1}(B_{i2}))]$. By Lemma \ref{lemmacongestion1},  $A_{i1}$ is optimal, and every $S_{i}$ satisfies all the conditions of the Modified Congestion Lemma. Therefore $EC_{g}(S_{i})$ is minimum.

For $1\leq j \leq \frac{k}{2}$, $E(C_{k}^{\star}(m))\setminus S_j$ has two components $B_{j1}$ and $B_{j2}$, where
\begin{eqnarray*}
V(B_{j1})=\left\{
\begin{array}{lcl}
\{k,k-1\} & ; & if\quad j=0 \\
\{j(k+1),j(k+1)+1\} & ; & if\quad j\neq0
\end{array}%
\right.
\end{eqnarray*}
Let $A_{i1}=G[V(g^{-1}(B_{i1}))]$
and $A_{i2}=G[V(g^{-1}(B_{i2}))]$. By Theorem \ref{lemmacongestion1},  $A_{j1}$, is optimal, and each $S_{j}$ satisfies tall the three conditions of the Modified Congestion Lemma. Therefore $EC_{g}(S_{j})$ is minimum.

For each $i$, $j$, $1 \leq i \leq \frac{k}{2}$, $1\leq j \leq\frac{m}{2}$, $E(C_{k}^{\star}(m))\backslash S_{i}^{j}$ has two components $B_{i1}^j$ and $B_{i2}^j$ where

\begin{equation*}
V(B_{i1}^j)=\left\{
\begin{array}{lcl}
\{j,j+1,\ldots,\frac{m}{2}+j-1\} \quad ;\quad if\quad i=0 \\
\{j((k+1)+i,j((k+1)+i+1\ldots,\frac{m}{2}+i+(k+2)j-1\} \quad ;\quad if\quad i\neq0
\end{array}%
\right.
\end{equation*}
Let $A_{i1}^j=G[V(g^{-1}(B_{i1}^j))]$
and $A_{i2}^j=G[V(g^{-1}(B_{i2}^j))]$. Since $A_{i1}^j$, is an optimal set, each  $S_{i}^{j}$ satisfies conditions $(i)$, $(ii)$ and $(iii)$ of the Modified Congestion Lemma. Therefore $EC_{g} (S_{i}^{j})$ is minimum. By Partition Lemma we can say that the wirelength is minimum.

\paragraph{\textbf{Case 2 ($m$ is odd):}}
For $i,j$, $1\leq i \leq m-1$, $1\leq j \leq m-1$, when $i$ and $j$ is odd, let $S_{i} =S_{j}$ be the edge cut that has all the edges of $C(i)$ together with the edge $B_c(i)$ and the edge between $B_c(i+\frac{k}{2})$ and $B_c(i+\frac{k}{2}+1)$. When $i$ and $j$ is even, let $S_{i} =S_{j}$ be the set of edges which has
edge of $B_c(i)$ together with the edge of $C(i+\frac{k}{2})$ and   $B_c(i+\frac{k}{2})$.
For $1 \leq i \leq m-1$, $1 \leq j \leq \frac{k}{2}$ let $S_{i}^{j}$ be the set of edges between $C(i)$ and $B_c(i)$.
Then $\{S_i,S_{j}:1\leq i\leq m-1\}\cup  \{ S_{i}^{j}:1 \leq i \leq m-1, 1\leq j \leq \frac{k}{2}\}$ is a partition of $[2E(C_{k}^{\star}(m))]$.

As in Case 1, it is easy to prove that the  wirelength is minimum.

\begin{theorem} \label{firstthm}
Let $A$ be the circulant graph $G(n;\pm \{1,2,\ldots,j\})$, $1\leq j < \left\lfloor n/2\right\rfloor$ and $B$ be the star of cycle $C_{k}^{\star}(m)$. Then then wirelength of embedding $A$ into $B$ is given by
\begin{equation*}
WL(A, B)=\left\{
\begin{array}{lcl}
\frac{k}{2}\!  \left\{\theta
_{A}   \left(\frac{m(k+1)}{2}\!\right)+\!\theta
_{A}(k)+\overset{m/2}{\underset{j=1}{\sum }} \theta_{A}(\frac{k}{2})\right\} & ; & if ~ m ~is~  even \\ \\

\frac{1}{2}\left\{\overset{m-1}{\underset{i=1}{\sum }} \theta_A\left(\frac{m(k+1)}{2}\right)+\overset{m-1}{\underset{j=1}{\sum }}\theta_A(k)+\overset{k/2}{\underset{j=1}{\sum }}\theta_A(\frac{k}{2})\right\}& ; & if ~m ~is ~odd.
\end{array}%
\right.
\end{equation*}

\noindent{\textit{Proof.}} Following the representation that are given in Embedding Algorithm B, we divide the proof into two cases.\\
\textbf{Case 1 ($m$ is even):}
By Modified Congestion Lemma,

\begin{enumerate}
\item[i)] $EC_g (S_i)= \theta_{A}\left(\frac{m(k+1)}{2}\right)$,       $1\leq i \leq \frac{k}{2}$
\item[ii)] $EC_g (S_{j})= \theta_{A}(k)$,       $1\leq j \leq \frac{k}{2}$
\item[iii)] $EC_f (S_{i}^{j})= \theta_{A}(\frac{k}{2}\!)$,      $1 \leq i \leq \frac{k}{2},$$1\leq j \leq\frac{m}{2}$.
\end{enumerate}
Then by Partition Lemma,
\begin{eqnarray*}
%\begin{split}
WL(G, C_{k}^{\star}(m)) &=& \bigg\{\overset{k/2}{\underset{i=1}{\sum }} EC_g (S_i)+\overset{k/2}{\underset{j=1}{\sum }} EC_g (S_{j})+\overset{m/2}{\underset{j=1}{\sum }} \overset{k/2}{\underset{i=1}{\sum }}EC_g (S_{i}^{j})\bigg\}\\
             &=& \bigg\{   \overset{k/2}{\underset{i=1}{\sum }} \theta_{A}\left(\frac{m(k+1)}{2}\!\right)+\overset{k/2}{\underset{i=1}{\sum }} \theta_{A}(k)+\overset{m/2}{\underset{j=1}{\sum }}\overset{k/2}{\underset{i=1}{\sum }} \theta_{A}\left(\frac{k}{2}\!\right)\bigg\}\\
             &=& \bigg\{ \frac{k}{2}   \theta_{A}\left(\frac{m(k+1)}{2}\!\right)+\frac{k}{2} \theta_{A}(k)+\frac{k}{2}\overset{m/2}{\underset{j=1}{\sum }} \theta_{A}\left(\frac{k}{2}\!\right)\bigg\}\\
             &=& \frac{k}{2}\!  \left\{\theta
_{A}   \left(\frac{m(k+1)}{2}\!\right)+\!\theta
_{A}(k)+\overset{m/2}{\underset{j=1}{\sum }} \theta_{A}\left(\frac{k}{2}\right)\right\}
%\end{split}
\end{eqnarray*}
\textbf{Case 2 ($m$  is Odd):}
By Modified Congestion Lemma,

\begin{enumerate}
\item[i)] $EC_g (S_i)= \theta_{A}\left(\frac{m(k+1)}{2}\right)$,     $1\leq i \leq m-1$
\item[ii\emph{})] $EC_g (S_{j})= \theta_{A}(k)$,       $1\leq j \leq m-1$
\item[iii)] $EC_g (S_{i}^{j})=\theta_{A}(\frac{k}{2})$,        $1 \leq i \leq m-1$, $1 \leq j \leq \frac{k}{2}$.
\end{enumerate}
Then by $2$-Partition Lemma,
\begin{eqnarray*}
%\begin{split}
WL(G, C_{k}^{\star}(m)) &=& \frac{1}{2}\! \left\{\overset{m-1}{\underset{i=1}{\sum }}EC_g (S_i)+\overset{m-1}{\underset{j=1}{\sum }} EC_g (S_{j})+\overset{m-1}{\underset{i=1}{\sum }}\overset{k/2}{\underset{j=1}{\sum }}  EC_g(S_{i}^{j})\right\}\\
        &=& \frac{1}{2}\left\{\overset{m-1}{\underset{i=1}{\sum }}\theta
           _{A}\left(\frac{m(k+1)}{2}\right)+\overset{m-1}{\underset{j=1}{\sum }} \theta_{A}(k)+\overset{m-1}{\underset{i=1}{\sum }}\overset{k/2}{\underset{j=1}{\sum }} \theta_{A}\left(\frac{k}{2}\right)\right\}\\
        &=& \frac{1}{2}\left\{\overset{m-1}{\underset{i=1}{\sum }} \theta_{A}\left(\frac{m(k+1)}{2}\right)+\overset{m-1}{\underset{j=1}{\sum }} \theta_{A}(k)+\overset{k/2}{\underset{j=1}{\sum }}\theta_A\left(\frac{k}{2}\right)\right\}.
%\end{split}
\end{eqnarray*}
\end{theorem}

\section{Implementation}

The embedding of the interconnection networks is known to study the interrelation among the graphs to find whether a specific guest graph is interrelated to the host graph. If a guest graph could effectively embedded into the host graph with minimum cost, then the technique obtained in the interconnection  network with a guest graph could be used in the interconnection network  with the host graph at less cost. 

\begin{figure}
\centering
\includegraphics[width=10 cm]{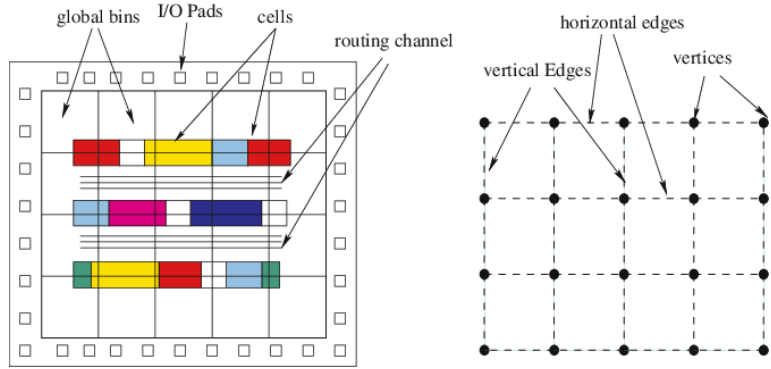}
%\vspace{-2 cm}
\caption{Graph representation of VLSI}
\label{fig2}
\end{figure}

The wirelength problem has been widely applied in different fields. One of the major application is the VLSI layout. Very Large Scale Integration (VLSI) concerns the process of integrating thousands of transistors into a single microchip to build an integrated circuit (IC). Though VLSI emerged in the 1970s, it keep remains as a strong methodological acquisition among the parallel and distributing computational models. It assists in the reduction of circuit size and increase of circuit operating speed by cost-effective and low power management. With the emergence of VLSI and its advantages, the number of applications of integrated circuits (ICs) in the general electronics, telecommunications, image processing and super computing systems have contributed to a very fast growing pace.

In VLSI layout, placing different processing elements play an influential role in reducing the communication overhead by minimizing the congestion of communication links between the processors. In general, any VLSI is comprised of a two step problem. To begin with, the placement problem is the process of placing the system’s processing elements in the most suitable position. Second, the routing problem is a method of reducing the wirelength and increasing the routability of a circuit system. To achieve this, IC with modules and wires in VLSI designs are transformed into a generalised graph with vertices representing the modules and edges representing the wires in such a way that total wirelength of connecting wires is minimized (see Figure.\ref{fig2}).

\begin{figure}
\centering
\includegraphics[width=8 cm]{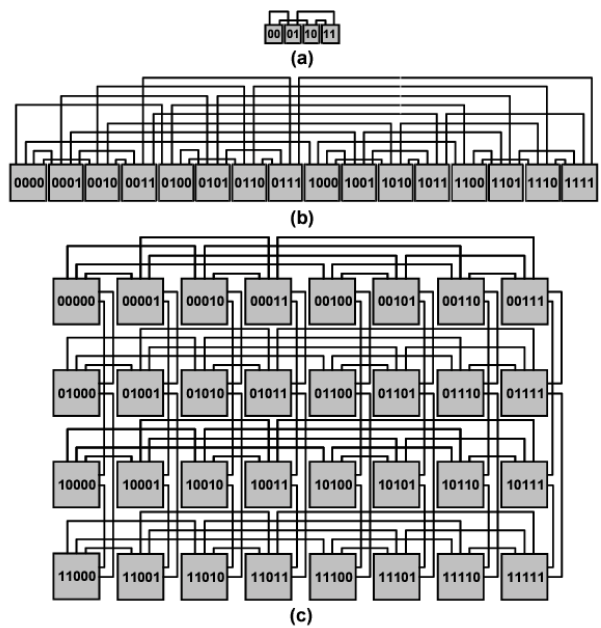}
%\vspace{-2 cm}
\caption{Multilayer VLSI Layout}
\label{fig4}
\end{figure}

The layout of interconnection networks has important cost and performance implications for single chip multiprocessors and parallel/distributed systems based on such components. Thus, there is currently renewed interest in finding efficient VLSI layouts for various interconnection networks. VLSI layout of interconnection networks is usually derived under the Thompson model, where two layers of wires are assumed. Many grid models such as the multilayer $2$-D grid and multilayer $3$-D grid models and triangular grid \cite{VrRs20} for VLSI layout of networks. In the multilayer grid model (See Figure \ref{fig4}), a network is viewed as a graph whose nodes correspond to processing elements and edges correspond to wires. Then the graph is embedded into many networks, including folded hypercubes, star graphs, star-connected cycles which lead to the following advantages such as the area of the layout can be reduced, the volume of the layout can be reduced, the maximum total length of wires along the routing path between any source-destination pair can be reduced. As the result the multilayer grid model for VLSI layout can also be embedded into cycle-of-ladders to minimize the wirelength.

\section{Concluding remarks}
The wirelength of embedding folded hypercube into grid has been obtained \cite{BeCh98}. In this paper, we have embedded circulant networks into star of cycle, and folded hypercube into cycle-of-ladders to find the minimum wirelength. Also the embedding constructed is simple, elegant and give exact wirelength. To obtain the wirelength of grid to cycle-of-ladders is under investigation.

%\nocite{*}
%\bibliographystyle{fundam}
%\bibliography{citations}

%%%%%%%%%%%%%%%%%%%%%%%%%%%%%%%%%%%%%%%%%%%%%%%%%%%%%%%%%%%%%%%%%%%%%%

\end{document}